**Babajide AFOLABI and Odile THIERY**
Laboratoire Lorraine de Recherche en Informatique et ses Applications (LORIA)
Campus Scientifique, BP 239 54506 Vandoeuvre - Lès - Nancy, France

# Using Users' Expectations to Adapt Business Intelligence Systems.

**Abstract:** This paper takes a look at the general characteristics of business or economic intelligence system. The role of the user within this type of system is emphasized. We propose two models which we consider important in order to adapt this system to the user. The first model is based on the definition of decisional problem and the second on the four cognitive phases of human learning. We also describe the application domain we are using to test these models in this type of system.

## 1. Introduction

### 1.1. Definitions: Business Intelligence and SIS

According to (Revelli, 1998), Business Intelligence (BI) "*is the process of collection, processing and diffusion of information that has as an objective, the reduction of uncertainty in the making of all strategic decisions*". This is referred to as Economic Intelligence (EI) to in this paper as this is the term used in France our country of research. Also, using EI avoids the confusion of limiting the process to the business world (Business Intelligence), since we think that every organisation/institution, be it socio-economic, political, cultural or otherwise can adopt the process). Also, we have adopted this term to avoid the erroneous thoughts that one has to be competitive (in Competitive Intelligence) in order to innovate.

The real interest of our research team is the aid that can be got from the use of the EI process in the resolution of decisional problems. The processes of collection, processing and diffusion can be fully or semi automated, we imagine this automation being powered by an Information System (IS). In actual fact, the EI process relies on the effective use IS. This type of information system belongs to the class of IS referred to as Strategic Information Systems (SIS), strategic in the sense that it contains information that is considered strategic because they are used in the decisional processes of the organisation (and not strategic because it is used to run the day to day activities of the organisation).

In its simplest form, a strategic information system (SIS) can be considered as an information system (IS) consisting of "*strategic information and permits the automation of the organisation to better satisfy the objectives of the management*". For instance, an IS that aids in the management of stocks, we denote this as SI-S. A SIS can also be seen as "*an IS that is dedicated to strategic decision making and contains only strategic type of information*". For example, an IS that permits the decision maker to observe sales by country for a number of years or that permits an information watcher to point up the choices made during the analysis of the result obtained from an information search on the web. This is denoted as S-IS. (Tardieu and Guthmann, 1991) (David and Thiery, 2003)

### 1.2. Economic Intelligence Systems (EIS)

The decisions taken, using an IS, are based on the information found in the IS and are also based on the user that has as an objective, the appropriation of such system for a decision making process. To us, an Economic Intelligence System EIS is a system that combines strategic information systems and user modelling domains. The final goal of a BIS is to help the user or the decision maker in his decision making process.

Figure 1 shows the architecture of an EIS as successive processes as proposed by the research team "SITE" (Modelling and Developing Economic Intelligence Systems of the

Lorraine Laboratory of IT Research and its Applications (LORIA) Nancy, France) one can easily identify the following four stages:

- **Selection:** selection which permits the constitution of the IS of the organisation that can be (i) the production database (that allows current usage of the organisation), (ii) all the information support for an information retrieval system (in documentation for an example) or (iii) a SIS based on a data warehouse. This information system is constituted from heterogeneous data and from heterogeneous sources with the aid of a filter.
- **Mapping:** mapping permits all users an access to the data in the IS. We are permitting two methods of access to the user: access by exploration and access by request. The exploration is based on a system of hypertexts. The requests are expressed with the aid of Boolean operators. The result of the mapping is a set of information.
- **Analysis:** in order to add value to the information found, techniques of analysis are applied on the results. For instance, the assistant of a head of department that we consider as the information watcher can present a summary of the results obtained on the information requested to his head of department.
- **Interpretation:** this means in general, the possibility of the user of the system being able to make the right decisions. It does not mean that the sole user of the system is the decision maker; it can include the information watcher. One can see then the interest in capturing the profile of the decision maker in a metadata stored on the data warehouse which can be used to build a specific data mart for a group of decision makers or even better a particular user.

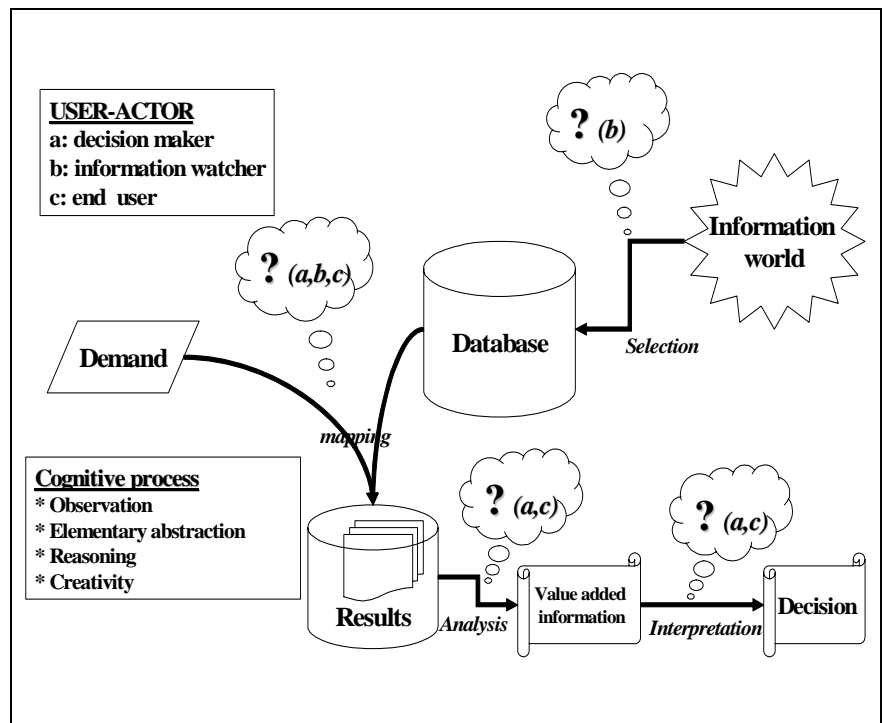

**Figure 1:** Architecture of an Economic Intelligence System.

Also in this process one can identify three main actors:

- **Decision maker:** this is the individual in the organization that is capable of identifying and posing a problem to be solved in terms of stake, risk or threat that weighs on the organization. In other words, he knows the needs of the organization, the stakes, the eventual risks and the threats the organization can be subjected to.
- **Information watcher:** this refers to the person within the organization that specializes in the methods of collection and analysis of information. His objective is to obtain indicators (using information) or value added information that the decision makers depend on for his decision process. After receiving the problem to be solved as expressed by the decision maker, the information watcher must translate it into information attributes to be collected and which are used to calculate the indicators.
- **End user:** this is the final user of the system; it can be either of the previously outlined users or neither of the two. This user is defined depending on which layer of the Economic intelligence system he interacts with.

Other works have shown that there may be other actors involved in this system (Knauf and David, 2004).

Earlier works by (Thiery and David, 2002) on personalization of responses in Information Retrieval Systems (IRS) adapted the four cognitive phases in the human learning process i.e.:

- **Observation phase:** here, the learner gathers information about his environment by observation.
- **Elementary abstraction phase:** the learner describes the objects observed using words, this corresponds to a phase of acquiring the vocabulary of the system being observed.
- **Reasoning and symbolization:** the learner starts to use the vocabulary acquired which implies a higher level of abstraction.
- **Creativity phase:** here the learner discovers and uses the knowledge that were not explicitly presented in the system.

## 2. Modelling the User

### 2.1. The role of the user

The user, who in this context can be the decision maker or in a larger context can any of the persons enumerated above, has a central role to play in an economic information system. His ability to efficiently use the system is directly proportional to his knowledge of the system. The first thing to do then will be to evaluate his knowledge of the system, use this knowledge to establish the importance of his role, his working habits, the most frequently used data etc.

Next, using this information, a personalised structure can be generated to improve his use of the system. A complete and robust work environment can enormously increase his efficiency. On the other hand, he can bring out the critical elements of the system, the errors, faults and missing points of the system. For a user – decision maker, the decision making process begins by acknowledging a decisional problem, that can be translated as a decisional need. The resolution of a decisional need consists in identifying the needs necessary for such resolution, be it informational, strategic, human etc. we will be concerned, at this time with only the need in information (informational needs). Informational need can be defined as a function of the user model, his environment and his objectives.

User modelling and adaptivity are needed to support:
- *Query adaptation:* the user's query may be adapted by the system to meet the user's specific needs as identified by the user model
- *Response adaptation:* the response of the system is based not just on any information but on information that relates to the user's goal or purpose.

### 2.2. Information need

The information need of a user is a concept that varies in definition, according to different researchers and according to the different users (Campbell and Rijsbergen, 1996), (Devadason and Pratap Lingam, 1996) and (Xie, 2000). We believe that the information need of a user is an informational representation of his decisional problem (Goria and Geffroy, 2004) and (Mizzaro, 1998). Defining a decisional problem implies certain level of knowledge on the user and his environment. Therefore, a decisional problem is a function of the user model, his environment and his objective.

We base our definition on that of (Bouaka and David, 2003) where a decisional problem was defined as

$P_{decisional}$ = f(Stake, Individual Characteristics, Environmental Parameters)

*Stake* (goal) is what the organization stands to loose or gain. It is made up of *Object, Signal and Hypothesis*. *Individual Characteristics* refer to the user, his behaviours and his preferences. This includes his *Cognitive Style,* his *Personality Traits*, and his *Identity*. *Environmental* parameters mean the input of the society on the organisation. This can be *Immediate* or *Global*.

### 2.3. The user's expectation

In the Economic Intelligence System, the process starts from the identification of the decisional problem. In order to resolve this problem it is translated into an informational need. The definition of informational need depends most of the time on the person or user involved his experiences, his functions, his environment etc as defined above. The user then forms his requests based on this informational need. The requests formed by the user are usually based on his expectations of the system and his idea that the system will respond to some of these expectations. We defined the user's expectations based on his model that as described above and the definition of his information need that is also dependent on his model. These expectations are contained in the potential knowledge field ("potential" because this may not actually be in use but it exists and is usable) of the organisation (figure 2).

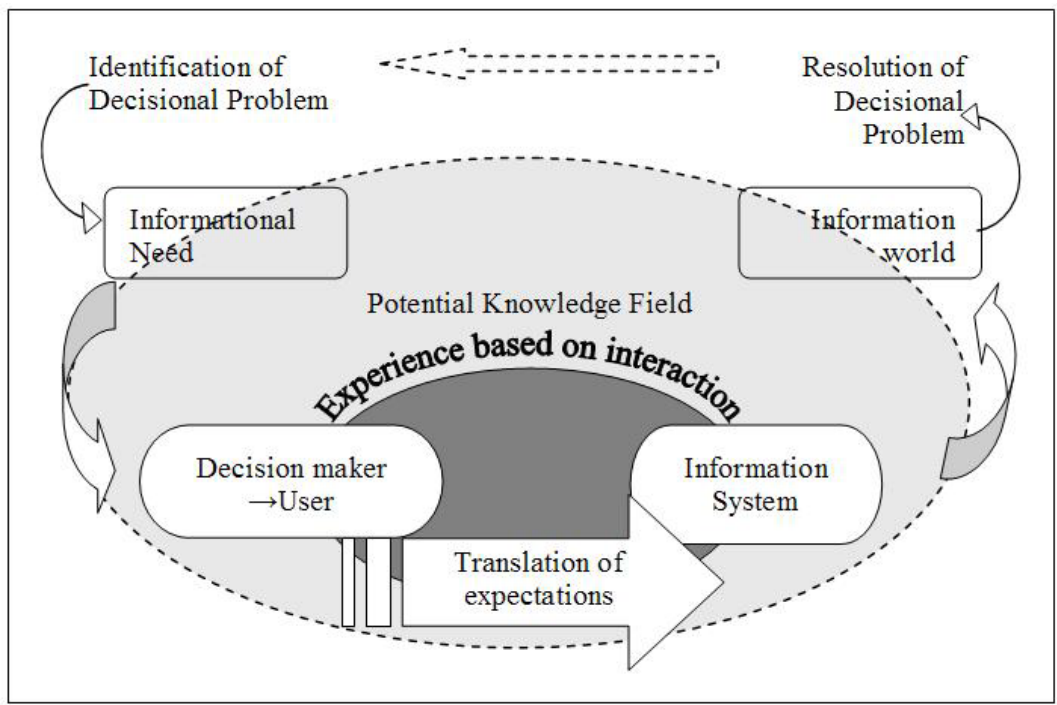

**Figure 2**: The use of user's expectation in an IS

The information system is alimented from an information world that is a global sum of all the credible sources available (within the organisation or external to it). The decisional problem can then be solved based on this available information. The resolution of a decisional problem can lead to the identification of another problem which means the system is in a continuous cycle. These expectations define some of his actions and these actions form the basis of his interactions with the system. Therefore we added these expectations in form of a variable called actions (activities) to the definition of decisional problem earlier cited. In other to complete this model for resolving decisional problems, we added the means that were implicated to achieve a resolution. Means in this case refer to the ways, methods and materials used.

Therefore the model for resolving decisional problems is based on:

$MP_{decisional}$ = f(Stake, Individual Characteristics, Environmental Parameters, Actions, Means)

### 2.4. The user model

Earlier models used in the Information System base of BIS were not complete since each user reacts differently according to his needs and his working habits, the possibilities of his evolution while using the system was not included. Also, a user/decision maker may have a need that is specific to him (in terms of his personality traits, cognitive style, preferences etc. as noted by Bouaka and David, 2003) that may not have been treated in the base. Our premier preoccupations include trying to complete a user model that will help the user in his evolutionary use of the system. This includes the system's ability to respond to his needs (informational) and allowing him to progress from a user learning to use the system to an expert user of the system without letting him feel he is going through the same processes over and over again.

The objective of user model is to be able to personalize the responses of the system. User modelling is the way a user and his behaviour are represented.

We transformed the four cognitive phases of human learning earlier mentioned into a user model in an IRS context. This is to give us a learning aspect to the system. The first two phases were compressed into **exploration** and this gives a model.

**M** = {Identity, Objective, {Activity} {Sub-sessions}}

Where:

Activity = {Activity-type, Classification, Evaluation}
Type = {Exploration, Request, Synthesis}
Classification = {Attributes, Constraints}
Evaluation = {System's solution, Degree of Pertinence}

- **Identity:** the identity of the user. This allows the individualisation of the historic of the sessions of the user.
- **Objective:** the principal objective or the real need of the user for the session.
- **Activity** (or actions in our model for resolving decisional problems)**:** A user activity that leads to the resolution of his information need. A session is composed of many activities and each activity is defined by three parameters: activity-type, classification, and evaluation.
- **Activity-type:** the types of activity correspond to the different phases of evocative user habits which are in this case exploration, request and synthesis.
- **Classification:** this is the approach we use to access stored information. The classification technique permits the user to express his information requirement in terms of the evocative phases that we are implementing. The user will be able to specify the attributes of the documents to classify and the constraints that are to be met by these documents.
- **Evaluation:** the user will be able to evaluate the pertinence of all the solutions proposed by the system. This evaluation relies on the degree of pertinence and the reasons for this judgement.
- **Sub-sessions:** a sub-session is represented exactly like a main session. The only difference is that the objective of the sub-session is associated to the objective of the main session and a sub-session will not constitute a session apart.

This user model permits the proposition of an information system architecture that relies on a cognitive user evolution. The user can: explore the information base to discover its contents; formulate his requests; add annotations; and link his information retrieval activities to a definite predetermined objective. The information on the user is updated with each use. In simpler words the user evolves with the system.

### 2.5. Collecting user's information

The user's information used in this IS are collected using the models described. It begins with the user entering the basic information concerning him (profile) explicitly. Firstly, we associate his requests or his information need(s) to the first model i.e. MP$_{decisional}$. This is used to improve his earlier given profile. Our hypothesis is that his use of the system improves with experience. His activities, called interactions, are stored as an experience base using the second model. This base is used to improve the model of the user and it helps follow the user with his use of the system. We used the experience of older users at a level of learning to guide new users at that same level or to kick start a new user expressing the same similarities. For instance a user that could not give attributes and the values associated with the attributes is considered as new to the system and will need to go through a system of observation to discover the attributes and the corresponding values.

## 3. Application domain

We are testing these methods, in the first instance, by applying the framework in information retrieval, using a base of documents published by researchers in a research centre. This base contains publications, historicized and grouped according to the habitual bibliographic nomenclature, of members of the research centre. We had worked on the classification, normalization and improvement of such electronic document resource and our objective is to constitute a real data warehouse of documents from which we could create all type of information analysis. In particular, we propose producing different data marts for the different group of users of the system, before going on to personalize the system to the individuals.

This presupposes that each of these groups of users has a different view of the data from the data warehouse and would want to propose to him the data that essentially respond to his needs.

Thus, while testing the earlier version, a user wanted to follow the evolution of publications of each research team. However, we found out that the attribute that could have helped in calculating this evolution was missing from the base. Another user wanted to know how represented African researchers are in terms of publications within this centre. The system could not provide the answer as nationalities of authors were not considered during the construction of the base. During the period of adaptation the information resources of the system is reengineered to contain the important attributes and or values that were missing which limit the responses got by the users.

## 4. Conclusion

If the user's expectations had been considered from the beginning, some of these questions would have been answered in the base. On verifying the identity of the user that wanted to follow the evolution of the publications of each research team we discover he is the director of the centre. If this user's information need had been properly defined, this information would have been taken care of.

The model for resolving decisional problems can be used along with the user model in an IRS context as defined above to further extract a lot of information on the user, his behaviours and why he behaves the way he does when in direct contact with the system.

Our next phase of the research is to construct a metadata from these two models which will serve as a tool in the re-conceptualisation the actual base.